# Computational Model of Motion Sickness Describing the Effects of Learning Exogenous Motion Dynamics

**Takahiro Wada[1]**

[1]College of Information Science and Engineering, Ritsumeikan University, Kusatsu, Shiga, Japan

**\* Correspondence:**
Takahiro Wada
twada@fc.ritsumei.ac.jp

## Abstract

The existing computational models used to estimate motion sickness are incapable of describing the fact that the predictability of motion patterns affects motion sickness. Therefore, the present study proposes a computational model to describe the effect of the predictability of dynamics or the pattern of motion stimuli on motion sickness. In the proposed model, a submodel—in which a recursive Gaussian process regression is used to represent human features of online learning and future prediction of motion dynamics—is combined with a conventional model of motion sickness based on an observer theory. A simulation experiment was conducted in which the proposed model predicted motion sickness caused by a 900 s horizontal movement. The movement was composed of a 9 m repetitive back-and-forth movement pattern with a pause. Regarding the motion condition, the direction and timing of the motion were varied as follows: a) Predictable motion (M_P): the direction of the motion and duration of the pause were set to 8 s; b) Motion with unpredicted direction (M_dU): the pause duration was fixed as in (P), but the motion direction was randomly determined; c) Motion with unpredicted timing (M_tU): the motion direction was fixed as in (M_P), but the pause duration was randomly selected from 4 to 12 s. The results obtained using the proposed model demonstrated that the predicted motion sickness incidence for (M_P) was smaller than those for (M_dU) and (M_tU) and no considerable difference was found between M_dU and M_tU. This tendency agrees with the sickness patterns observed in a previous experimental study in which the human participants were subject to motion conditions similar to those used in our simulations. Moreover, no significant differences were found in the predicted motion sickness incidences at different conditions when the conventional model was used.

## 1    Introduction

Motion sickness is caused by various types of body movements and visual information and can be experienced in our daily lives, such as in cars, ships, airplanes, and virtual environments. Owing to the recent advancements in vehicular control and mobile computing technology, the motion and visual stimuli received by humans have diversified, increasing concerns about motion sickness.

There are multiple hypotheses on the mechanism of motion sickness (Shupak and Gordon, 2006). The sensory conflict or neural mismatch theories that postulate that discrepancies existing among the multiple sources of sensory signals or between the sensory information and anticipatory information generated by the central nervous system (CNS) lead to motion sickness (Reason, 1978). Oman (1991) stated that the anticipatory signal may arise from the internal model that is thought to be built within our CNS. In the subjective vertical conflict (SVC) theory, which is a branch of the sensory conflict theory, it is postulated that conflicts between the vertical direction of the body given by the sensory signals and those estimated by our CNS or by the internal model lead to motion sickness (Bles et al., 1998). In addition, there exist several types of possible hypotheses for the etiological mechanisms of motion sickness. Riccio and Stoffregen (1991) postulated that motion sickness can be explained as instability of body posture control. Ebenholtz et al. (1994) hypothesized that motion sickness can be understood as a result of the eye movements controlled by the vestibular nuclei. For other hypotheses, please refer to the review article by Shupak and Gordon (2006).

Computational or quantitative understanding of the sensory conflict theory and the SVC theory is progressing. Pioneering work was conducted by Oman, who proposed a computational model of the sensory conflict theory in which the internal model in our CNS is modeled using the observer framework of the modern control theory (Oman, 1982). Following the publication of Bos and Bles (1998), several computational models have been developed based on the SVC hypothesis. The pioneering work on the computational model of the SVC hypothesis calculated the motion sickness incidence (MSI) defined as the percentage of the people who would vomit with given motion stimuli for the vertical motion acceleration of the head. Kamiji et al. (2007) and Wada et al. (2018) expanded this to six degree-of-freedom (DOF) motion of the head, with 3DOF translational head acceleration and 3DOF head angular velocity based on the introduction of the dynamics of the otolith–canal interaction. Both of them estimated the MSI from vestibular or physical head motion patterns. Computational models of the SVC hypothesis that can deal with visual motion stimulus have been developed, including a 1DOF model (Braccesi and Cianetti, 2011) as the expansion of that proposed by Bos and Bles (1998), and a 6DOF model (Wada et al., 2020) as the expansion of those proposed by Kamiji et al. (2007) and Wada et al. (2018).

Various publications have demonstrated that the simplicity or predictability of the motion stimulus affects the severity of motion sickness. For example, it has been suggested that the severity of the sickness differs depending on whether the motion stimulus is a simple sine wave or a combination of multiple sine waves (Guignard and McCauley, 1982). More recently, an experiment in which the participants rode in an experimental cart without an external view and moved in the horizontal direction demonstrated that the sickness was significantly less profound when the motion pattern was predictable compared with the case in which the motion direction and timing were random (Kuiper et al., 2019). An experiment that used the same apparatus and similar motion patterns showed that even when the direction and timing of the motion were randomly determined, the presentation of audio cues that determined the direction and timing of the motion reduced motion sickness (Kuiper et al., 2020). Furthermore, it was demonstrated that a human machine interface that tells car occupants when and which way to turn reduced motion sickness (Karjantao et al., 2018). However, no model has been proposed that can qualitatively and quantitatively explain the effects of such future motion prediction on motion sickness.

The purpose of the present study is to propose a computational model that can predict the effect of the predictability of the dynamics or the pattern of the motion stimulus on motion sickness by expanding the conventional model of motion sickness. The validity of the model is examined by comparing the computational results with experimental results obtained when participants were subjected to similar patterns of motion stimuli.

## 2    Modeling Effects of Motion Prediction based on Sensory Conflict Theory

### 2.1    Model of Sensory Conflict Theory based on Observer Framework

Reason (1978) emphasized that the sources of conflict in the sensory conflict theory or neural mismatch theory are thought to be associated with the discrepancy between the sensory afferent signal and the expected signal calculated by the CNS rather than with simple comparison between sensory modalities, such as visual and vestibular sensations. The key component used to calculate the expected afferent is a "neural store," which retains the exposure history.

Reason (1978) stated that the neural store

[A] Retains the history of the combinations of efferent and the resultant sensory afferent signals in the case of self-produced movement and calculates the expected afferent signal from a given efferent signal and

[B] Is involved with the generation of the expected afferent signal even for passive movements when the motion disturbances can be predicted.

In pioneering work, Oman et al. (1982, 1991) successfully built a computational model of the neural mismatch theory from the viewpoint of [A] with the introduction of the observer approach (Fig. 1). In this model, the expected sensory afferents were calculated from the efferent copy and other signals with the use of internal models of the body and sensory dynamics that are thought to be built in the CNS. The difference between the expected and actual sensory afferents is considered to be the sensory conflict that leads to symptoms of motion sickness. The sensory conflict is fed back to the internal model to improve the accuracy of the estimated motion, which is used for motion control. This process is expected to reduce sensory conflict. When the conflict cannot be eliminated by the presence of exogenous motion disturbance, which is defined as a motion caused by external forces, the sickness is exacerbated. Several computational models of sensory conflict or neural mismatch theories proposed thus far can be interpreted based on this idea (Bos and Bles, 1998; Kamiji et al., 2007).

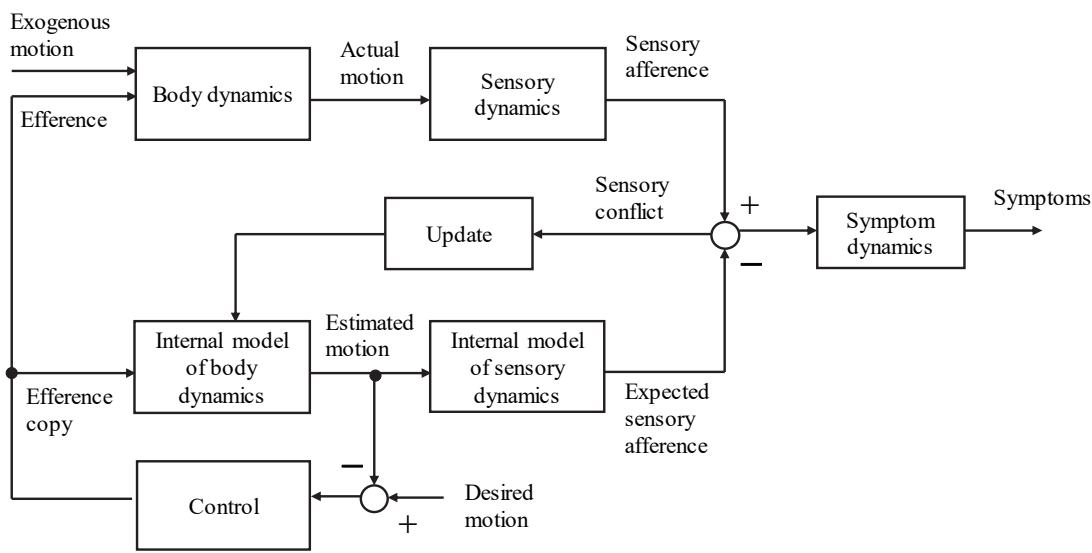

Figure 1. Schematic block diagram of sensory conflict theory based on internal model hypothesis.

## 2.2 Effect of Motion Prediction on Motion Sickness

Prior publications demonstrated that motion prediction reduces motion sickness. From the findings thus far, predictions that facilitate the reduction of motion sickness are believed to be obtained from

[A] Efference copy in active motion (Oman, 1991; Rolnick and Lubow, 1991; Wada et al., 2018) and

[B] Knowledge or information about the exogenous motion disturbances that can be obtained by

(i) Learning the dynamics when the movement is periodic (Kuiper et al., 2019),

(ii) Visual information (Turner and Griffin, 1999; Butler and Griffin, 2006; Wada and Yoshida, 2016), and

(iii) Other cues, such as artificial signals (Karjanto et al., 2018; Kuiper et al. 2020).

The efference copy [A] is shown in Fig. 1. The knowledge or information about the exogenous motion disturbances [B] is also believed to be used to increase the accuracy of the self-motion perception. Thus, we hypothesized that the effect of knowledge about exogenous motion [B] can be described in a similar way as the efference copy shown in Fig. 2. To the best of the author's knowledge, to date, there are no computational or conceptual models that deal systematically with the effects of the knowledge of exogenous motion disturbances.

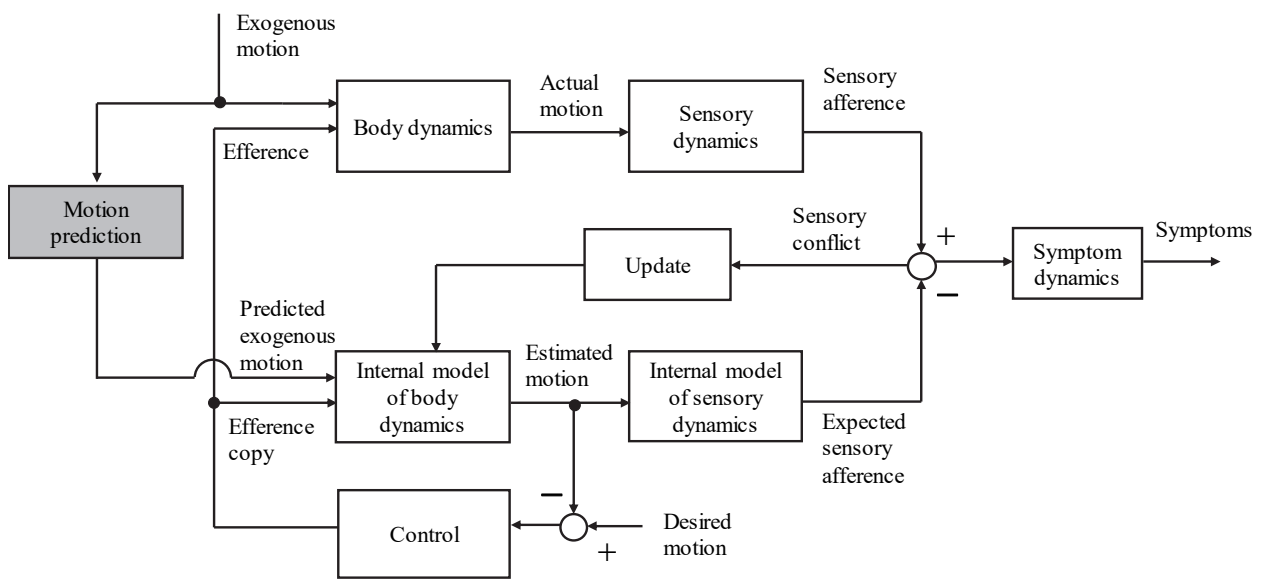

Figure 2. Schematic block diagram of sensory conflict theory capable of capturing the effect of human prediction of exogenous motion disturbance.

## 3 Computational Model of SVC Theory of Motion Sickness Including the Effects of Motion Dynamics Learning and Motion Anticipation

The present study proposes a concrete computational model of motion sickness that can describe the effect of the predictability of the exogenous motion for (i) periodic motions.

The function of learning and predicting the dynamics of exogenous disturbances for (i) periodic motion is incorporated into a motion sickness model built based on human motion perception and the observer approach, as shown in Fig. 2. For simplicity, when a concrete quantitative model is built, only situations in which the body motion is caused by external forces, i.e., passive motion, is considered.

## 3.1 Formulation of the Learning Dynamics of Exogenous Disturbance and Motion Prediction

For simplicity, let us consider one-dimensional translational human motion. Assume an environment exposed to acceleration, meaning that the exogenous motion that should be learnt and predicted is the body acceleration caused by external force. The motion from the past and the motion at the next time step are expressed by a nonlinear autoregressive (NAR) model described by Eq. (1):

$$y_i = f_i(\boldsymbol{x}_i) + w_i,$$

(1)

where $y_i \coloneqq \alpha_{i+1} \in R$, $\boldsymbol{x}_i \coloneqq [\alpha_i, \alpha_{i-1}, \alpha_{i-2}, \cdots, \alpha_{i-d-1}]^T \in R^d$ and $w_i$ denotes noise. In the case that a certain pattern can be found in the motion, that is, when Eq. (1) is rewritten in the form of Eq. (2), it is considered that a human can learn this pattern $f(\cdot)$ sequentially and can then use it to predict the subsequent motion.

$$y_i = f(\boldsymbol{x}_i) + w_i$$

(2)

A recursive Gaussian process regression model (Huber, 2014), one of the Bayesian, nonlinear, nonparametric regression models, was used in the present study to model the process of sequential learning for various motion patterns (see Eq. (2)) and to predict the subsequent motion in the presence of uncertainty.

## 3.2 Recursive Gaussian Process Regression Model for Sequential Learning

Let us consider constructing a nonlinear regression model using the function in Eq. (2) with the Gaussian process regression (GPR) method (Williams and Rasumussen, 1996) and its family that can learn nonlinear functions nonparametrically. The noise $w_i \in R^1$ in Eq. (2) is assumed to be normally distributed $p(w_i) = N(0, \sigma_w^2)$. In the GPR, the function $f(\cdot)$ of Eq. (2) is modeled from the location vectors $X \coloneqq [\boldsymbol{x}_1, \boldsymbol{x}_2, \dots, \boldsymbol{x}_m] \in R^{d \times m}$ according to the following Gaussian distribution:

$$p(f \mid X) = N(f \mid 0, K(X, X)),$$

(3)

where the covariance matrix is defined as follows:

$$K(X, X) \coloneqq [k(\boldsymbol{x}_i, \boldsymbol{x}_j)]_{i,j} \in R^{m \times m}.$$

(4)

For a kernel function $k(\boldsymbol{x}_i, \boldsymbol{x}_j)$, the squared exponential kernel is used as follows:

$$k(\boldsymbol{x}_i, \boldsymbol{x}_j) := \alpha^2 \exp\left\{-\frac{1}{2}(\boldsymbol{x}_i - \boldsymbol{x}_j)^T \Lambda^{-1}(\boldsymbol{x}_i - \boldsymbol{x}_j)\right\} \in R^1, \quad (5)$$

where various types of symmetric positive definite matrices can be used as the kernel. Parameters $\alpha$, $\Lambda = diag[\lambda_i] \in R^{d \times d}$, and $\sigma_w$, which denotes the standard deviation of the noise $w_i$ in Eq. (5), are called hyperparameters.

In the GPR (Williams and Rasmussen, 1996), the training dataset $D := \{(\boldsymbol{x}_1, y_1), (\boldsymbol{x}_2, y_2), ..., (\boldsymbol{x}_n, y_n)\}$ pairs of the input and output of Eq. (2) is provided at one time instant, and model learning is performed only once. In contrast, a recursive Gaussian process regression (RGPR) method was proposed by Huber (2014), wherein when the input $\boldsymbol{x}_t$ is given at each time $t$, the mean $\mu_t^p$ and variance $C_t^p$ of the corresponding output are predicted, and the model is updated based on the observed data $y_t$ of the obtained output given that it is suitable for the modeling of the function of sequential learning and the prediction of the motion. In the RGPR, a matrix composed of $n$ location vectors $X := [\boldsymbol{x}_1, \boldsymbol{x}_2, ..., \boldsymbol{x}_n] \in R^{d \times n}$ is provided in advance instead of dataset $D$.

The following is the algorithm of the RGPR model (Huber, 2014) used in the present study.

[Initialization]

Prepare location vectors $X := [\boldsymbol{x}_1, \boldsymbol{x}_2, ..., \boldsymbol{x}_m]$ $(\in R^{d \times m})$. In this study, $X$ was determined to select m vectors from exogenous motion vector data based on the publication by Snelson and Ghahramani (2006).

At $t = 0$, calculate $\boldsymbol{\mu}_0^f = \boldsymbol{0}$, $C_0^f = K(X, X)$.

[Inference step] At time step $t = t$, given the input data $\boldsymbol{x}_t$,

1) Calculate the gain matrix $J_t$ according to Eq. (6):

$$J_t := \boldsymbol{k}(\boldsymbol{x}_t, X) K(X, X)^{-1} \quad (\in R^{1 \times m}). \quad (6)$$

2) Calculate the mean $\mu_t^p$ with Eq. (7) and covariance matrix $C_t^p$ using Eq. (8):

$$\mu_t^p := J_t \boldsymbol{\mu}_{t-1}^f \quad (\in R^1) \quad (7)$$

$$C_t^p := k(\boldsymbol{x}_t, \boldsymbol{x}_t) - J_t \{\boldsymbol{k}(X, \boldsymbol{x}_t) - C_{t-1}^f J_t^T\} \quad (\in R^1). \quad (8)$$

[Update step] Given observation data $y_t$ which corresponds to the input $\boldsymbol{x}_t$,

3) Calculate the gain matrix $\tilde{G}_t$ with Eq. (9):

$$\tilde{G}_t := C_{t-1}^f J_t^T (C_t^p + \sigma^2)^{-1} \quad (\in R^{m \times 1}). \quad (9)$$

4) Update the mean $\boldsymbol{\mu}_t^f$ and covariance matrix $C_t^f$ of the regression function with Eqs. (10) and (11), respectively,

$$\boldsymbol{\mu}_t^f := \boldsymbol{\mu}_{t-1}^f + \tilde{G}_t(y_t - \mu_t^p) \quad (\in R^{m \times 1}) \tag{10}$$

$$C_t^f := C_{t-1}^f - \tilde{G}_t J_t C_{t-1}^f = (I - \tilde{G}_t J_t)C_{t-1}^f \quad (\in R^{m \times m}) . \tag{11}$$

The kernel functions and their associated matrices and vectors that appear above are defined below.

$$k(\boldsymbol{x}_i, \boldsymbol{x}_j) := \alpha^2 \exp\left\{-\frac{1}{2}(\boldsymbol{x}_i - \boldsymbol{x}_j)^T \Lambda^{-1}(\boldsymbol{x}_i - \boldsymbol{x}_j)\right\} \quad (\in R^1) \tag{12}$$

$$K(X, X) := [k(\boldsymbol{x}_i, \boldsymbol{x}_j)]_{i,j} \qquad (\in R^{m \times m}) \tag{13}$$

$$\boldsymbol{k}(X, \boldsymbol{x}_i) := col[k(\boldsymbol{x}_i, \boldsymbol{x}_i))]_i \quad (\in R^{m \times 1}) \tag{14}$$

$$\boldsymbol{k}(\boldsymbol{x}_i, X) := row[k(\boldsymbol{x}_i, \boldsymbol{x}_i))]_i = \boldsymbol{k}^T(X, \boldsymbol{x}_i) \quad (\in R^{m \times 1}) \tag{15}$$

### 3.3 Computational Model of Motion Sickness with Effects of Learning and Prediction of Exogenous Disturbance

We focus herein on the SVC hypothesis (Bles et al., 1998), which constitutes the most computationally modeled hypothesis among all the motion sickness hypotheses. Let us consider adding the effects of learning the dynamics of exogenous disturbances and predicting the next motion introduced in the previous section. Bos and Bles (1998) developed a computational model of the SVC hypothesis (hereafter referred to as 1DOF-SVC) for translational motion in the vertical direction with 1DOF. This model has been expanded into a computational model of 6DOF motion with rotational motion (hereafter referred to as the 6DOF-SVC model) and a block related to the effect of the accuracy of self-motion perception and efference copy has been introduced to describe the difference in motion sickness between active and passive motion (Kamiji et al., 2007; Wada et al., 2018). By inputting zero into the angular velocity vector of the 6DOF-SVC model, an expansion of the 1DOF-SVC model of Bos and Bles (1998) into a three-dimensional translational model with an efference copy block (hereafter referred to as the 3DOF-SVC model) was obtained.

The present study proposes a computational model of motion sickness by adding the parts related to the effects of learning and prediction of the exogenous motion patterns. This model is based on the method described in the previous section on the 3DOF-SVC model (Fig. 3).

The inputs to the model are the gravito-inertial acceleration (GIA) $\boldsymbol{\alpha}$ ($\in R^3$), which is defined as

$$\boldsymbol{\alpha} := \boldsymbol{a} + \boldsymbol{g} , \tag{12}$$

where $\boldsymbol{g}$ is the acceleration due to gravity and a is the inertial acceleration. The vector $\boldsymbol{\alpha}$ is input to the otolith and is represented by the block marked OTO in Fig. 3, which represents otolith. The transfer function of the OTO is given by a unit matrix. The sensed vertical direction in the head-fixed frame $\boldsymbol{v}$ is estimated by block LP in Fig. 3 and is defined as the following lowpass filter:

$$\frac{d\boldsymbol{v}}{dt} = \frac{1}{\tau}(\boldsymbol{\alpha} - \boldsymbol{v}), \tag{13}$$

where the time constant is $\tau = 2$ s (Correia Gracio et al., 2013). In the figure, the lower part of the block diagram shows an internal model that is believed to exist in the CNS. Blocks $\overline{\text{OTO}}$ denote the internal models of OTO. The transfer function of $\overline{\text{OTO}}$ is given by a unit matrix. The internal model of LP is illustrated as $\overline{\text{LP}}$ and is assumed to be identical to LP. The internal model outputs the expected afferent signal $\hat{\boldsymbol{v}}$. The vector $\Delta\boldsymbol{v}$ denotes the error between the sensory afferent and the expected one. The error is used to calculate the predicted MSI through the Hill function $(\|\Delta\boldsymbol{v}\|/b)^2 / \{1 + (\|\Delta\boldsymbol{v}\|/b)^2\}$, which models the nonlinear relationship between the MSI and the magnitude of the vertical conflict and the second-order lag with a large time constant $P/(\tau_l s + 1)^2$, as depicted in Fig. 3 (Bos and Bles, 1998). This error $\Delta\boldsymbol{v}$ is fed back to the internal model through integration and gain $K$ to decrease the error.

The exogenous motion learning and prediction part on the left side is newly added in the present study. The input into the online learning and prediction part is

$$\hat{\boldsymbol{\alpha}} := \boldsymbol{\alpha} + \boldsymbol{n}, \tag{14}$$

where $\boldsymbol{n}$ denotes noise that is assumed to be normally distributed $N(0, 1 \times 10^{-4})$. The input variable vector for the nonlinear autoregressive model of Eq. (1) is defined as $\boldsymbol{x}_t := [\hat{\alpha}_{t-1}, \hat{\alpha}_{t-2}, \cdots, \hat{\alpha}_{t-d-1}]^T \in R^d$, where $d = T_w / \Delta t$, the time window is $T_w = 10$ [s], and the sampling time is $\Delta t = 0.01$ [s]. The GIA mean $\mu_t^p$ and variance $C_t^p$ in the next time step $t$ can be predicted by the RGPR model. The RGPR model is updated sequentially using the observed data $\hat{\alpha}_t$ based on the algorithm introduced in the previous section.

The predicted exogenous motion at time $t$, $\tilde{\alpha}_t$ is determined by Eq. (15)

$$\tilde{\alpha}_t := k_{pr} \mu_t^p, \tag{15}$$

where the weight $k_{pr}$ is determined by Eq. (16)

$$k_{pr} := c_1 \exp(-c_2 C_t^p), \tag{16}$$

based on an assumption that the weight decreases when the confidence of the prediction is low according to the analogy with the multisensory integration (Ernst, M.O., and Bulthoff, 2004). The coefficients are determined as $c_1 = 0.8, c_2 = 3$ by trial and error. The predicted exogenous motion $\tilde{\alpha}$ together with the feedback signal from the vertical conflict is input into the internal model. The accurate prediction with high confidence is expected to reduce the conflict and resultant MSI.

The model explained above can be formally expressed by the following discrete-time state and output equations:

$$\boldsymbol{\xi}_{t+1} = \boldsymbol{F}(\boldsymbol{\xi}_t) + \boldsymbol{\alpha}_t + \tilde{\boldsymbol{\alpha}}_t \tag{17}$$

$$m_t = \boldsymbol{c}^T \boldsymbol{\xi}_t, \tag{18}$$

where the inputs are the GIA vector $\boldsymbol{\alpha}_t$ ($\in R^3$) and its estimate $\tilde{\boldsymbol{\alpha}}_t$ ($\in R^3$) according to the learning and prediction parts, and $\boldsymbol{\xi}_t := [\boldsymbol{v}_t^T, \hat{\boldsymbol{v}}_t^T, \hat{\boldsymbol{\alpha}}_t^T, \dot{m}_t, m_t]^T$ ($\in R^{11}$) is a state vector. Scalar $m_t$ denotes the predicted MSI as the output of the model, and $\boldsymbol{c} := [\boldsymbol{0}_{1 \times 10}, 1]^T$.

All parameters used in the model are listed in Table 1.

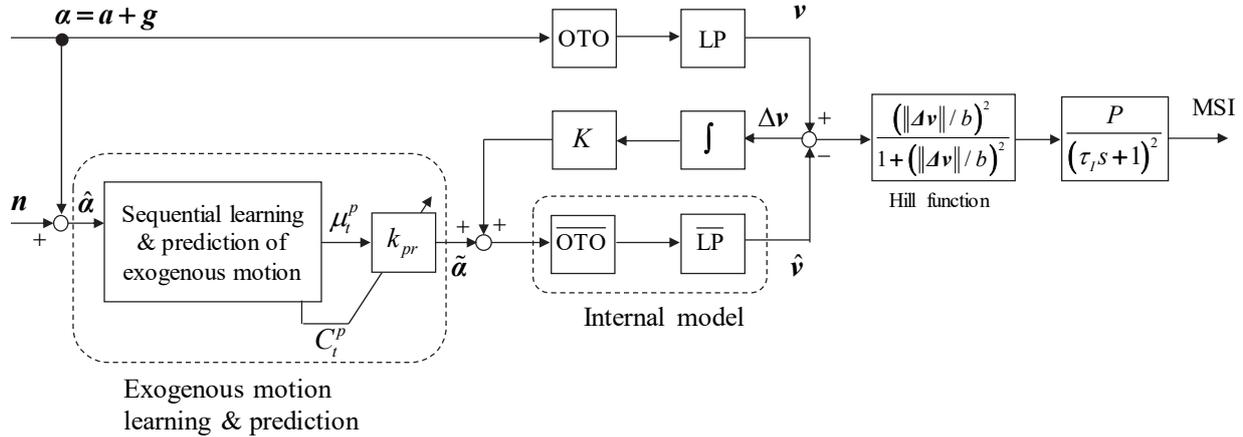

Figure 3. Block diagram of sensory conflict theory based on the internal model hypothesis.

## 4 Simulation Experiments

### 4.1 Motion Profile Condition

In an experiment conducted by Kuiper et al. (2019), participants on a cart (in which external vision and airflow cues were blocked) were moved back and forth for 900 s. In the experiments, the effects of predictability of the motion patterns on motion sickness were investigated. In the present study, we adopted almost the same motion pattern conditions as those in the study by Kuiper et al. (2019).

The following three levels were set as the motion pattern factor. All three motion profiles were constructed as repetitive, single forward and backward motions, while the predictability of the motion direction and time interval of the repetition were changed in different conditions.

1) Predictable (M_P) condition

The direction of motion and the time interval (8 s) were both constant.

2) Unpredictable direction (M_dU) condition

The direction of motion was randomly selected to be either the same or opposite to the M_P condition. The time interval was constant for 8 s, as in the M_P condition.

3) Unpredictable timing (M_tU) condition

The time interval was randomly selected from a uniform distribution from 4 to 12 s. The direction of motion was constant, as it was in the case of the M_P condition.

Fig. 4 shows the longitudinal displacement used in these calculations. Every displacement lasted for 8 s with the peak acceleration set to 2.5 m/s². The resultant amplitude was approximately 9.0 m. Note that the root-mean-square values of the accelerations among different conditions were identical.

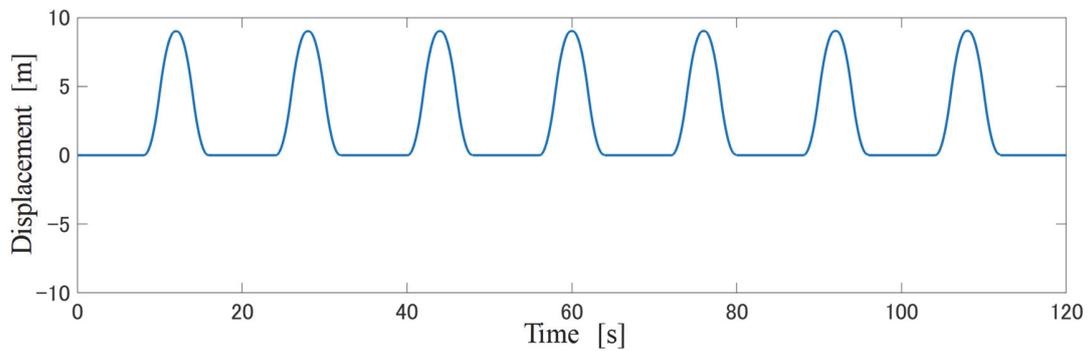

(A) M_P condition

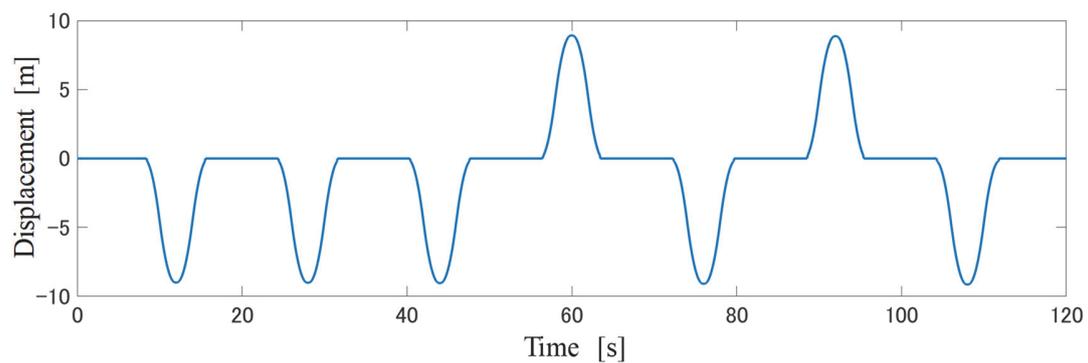

(B) M_dU condition

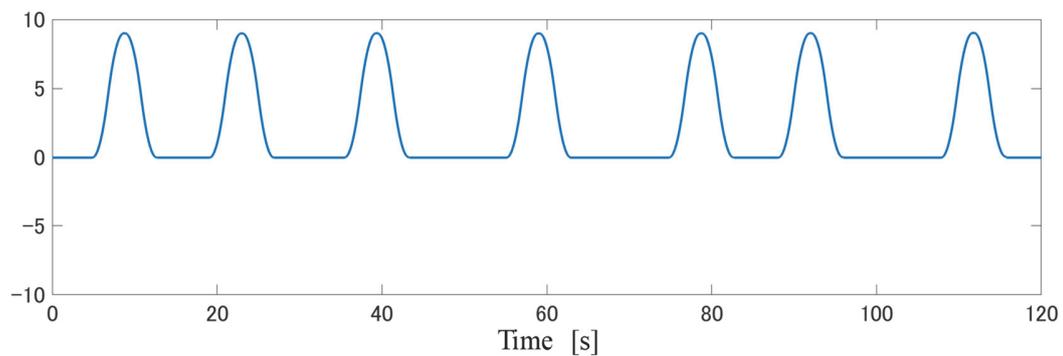

(C) M_tU condition

Figure 4. Motion profiles for three types of motion in the first 120 s.

## 4.2 Calculation Conditions

The following three levels were set as the calculation conditions to confirm the validity of the method.

(I) Control (C_Control) condition

In this condition, motion pattern prediction was not used. We used $\tilde{\alpha}_t^i := k_{pr}\alpha_t^i$ instead of $\tilde{\alpha}_t^i := k_{pr}\mu_t^p$, and $k_{pr} = 0.5$ was used according to Wada et al. (2018).

(II) Learning and prediction based on GPR (C_GPR) condition

The proposed methods described in section 3.3, including leaning and predicting motion patterns using GPR and gain tuning according to $C_t^p$, were employed.

## 5 Results

### 5.1 Calculated Results Based on the Proposed Model

First, Fig. 5 shows the calculated results of MSI subject to the C_GPR condition according to the proposed method. In the M_P condition of the motion profile, which was the periodic motion profile, high-prediction accuracy was achieved, as evidenced by the small error between the mean acceleration predicted by the motion learning and prediction part, and the actual acceleration. The variance predicted by the motion learning and prediction part converged to value of <0.3 within 10 s. As a result, the gain $k_{pr}$ assumed a large value, which was approximately equal to 0.6 except when the variance decreased, leading to an increase in the accuracy of self-motion perception and smaller predicted MSI compared with those in the other two motion profile conditions. In the M_dU and M_tUconditions, the error between the mean acceleration predicted by the motion learning and prediction part and the actual motion profile increased on a sporadic basis. The variance increased sporadically in the M_dU and M_tU conditions and resulted in a smaller gain $k_{pr}$ than that achieved in the M_P condition. These led to a larger predicted MSI than that in the M_P condition. As a result, motion learning and prediction did not facilitate an increase in the accuracy of self-motion perception in the M_dU and M_tU conditions. Consequently, the predicted MSI was the smallest for the M_P condition, and no considerable difference was found between M_dU and M_tU.

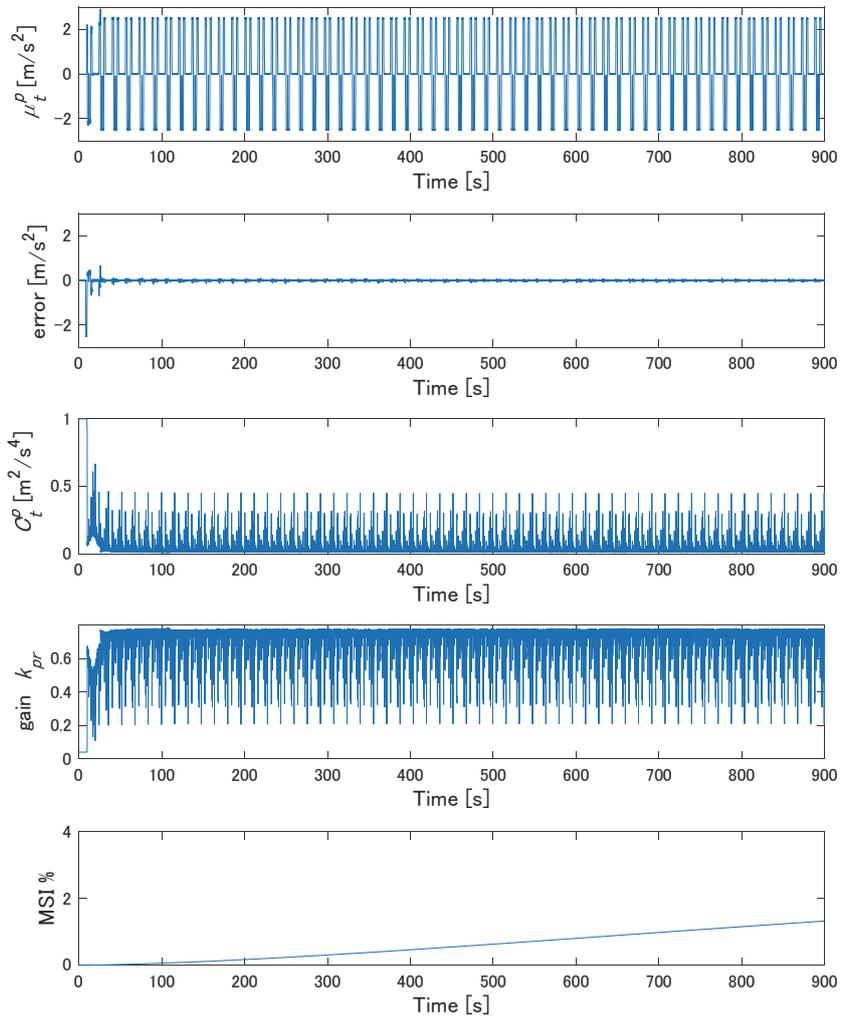

(A) M_P condition

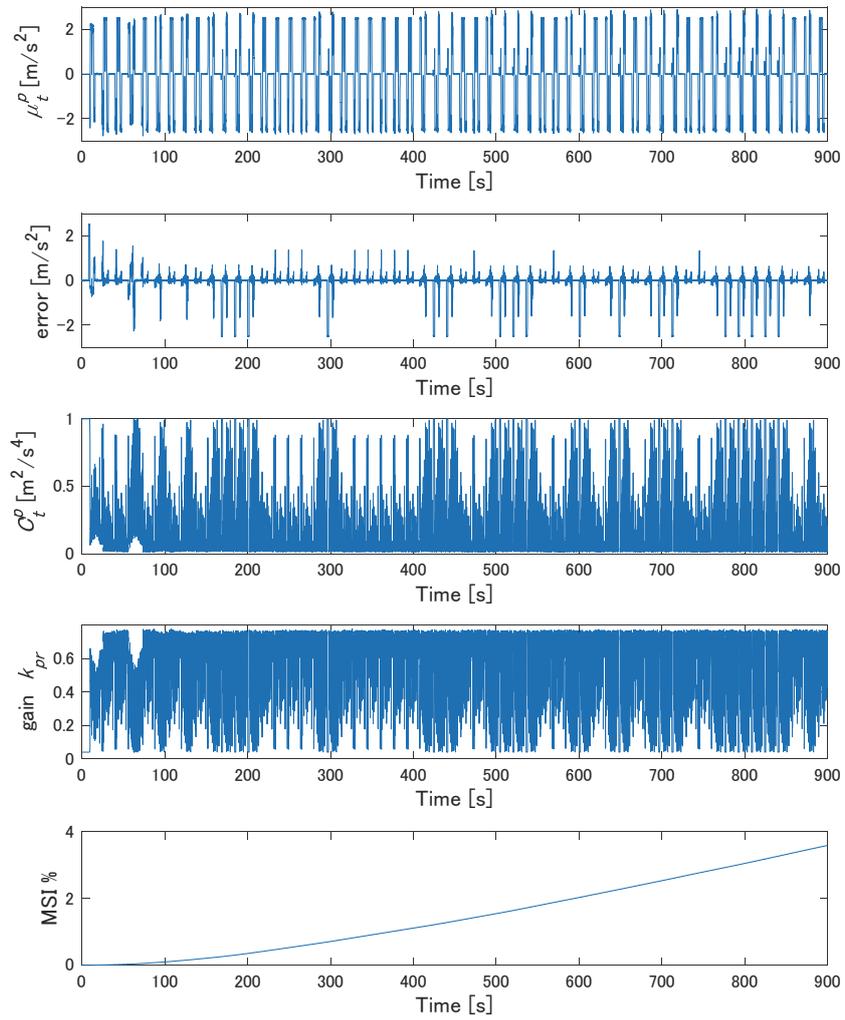

(B) M_dU condition

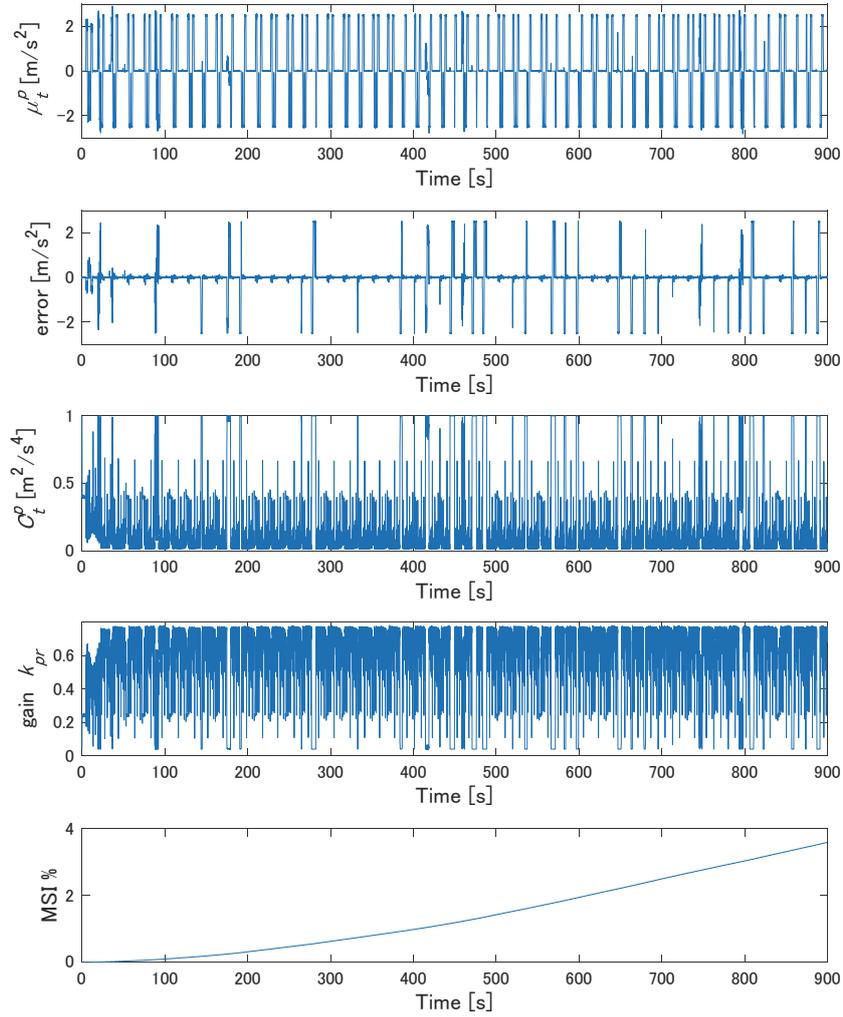

(C) M_tU condition

Figure 5. Temporal histories of representative variables with the proposed model for every motion profile condition: (i) M_P, (ii) M_dU, and (iii) M_tU. The scalars $\mu_t^p$ and $C_t^p$ denote the mean and variance of acceleration predicted by the proposed method, respectively. The error depicted on the second row was calculated as $\mu_t^p - \alpha_t$: the discrepancy between the predicted mean and the given acceleration of the exogenous disturbance. The gain $k_{pr}$ was determined by Eq. (16) according to $C_t^p$. The predicted MSI by the proposed model is shown as the output of the model.

## 5.2 Comparison of the Predicted MSIs in Different Conditions

Fig. 6 shows the predicted MSI at $t = 900$ s at the end of the exogenous motion. Under the C_Control condition, no significant differences in the predicted MSIs were observed among motion profile conditions M_P, M_dU, and M_tU. Under the C_GPR condition, the MSI at the M_P condition was significantly smaller than those at the M_dU and M_tU conditions. No significant difference was found between the M_dU and M_tU conditions.

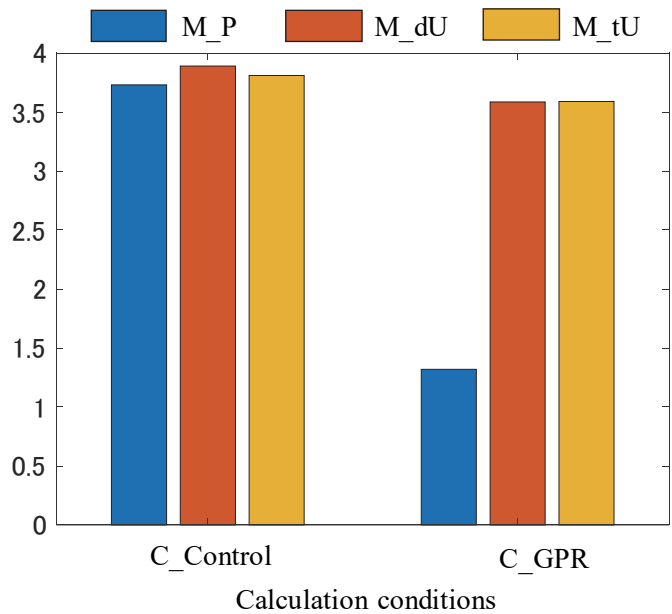

Figure 6. Predicted motion sickness incidence under different conditions.

## 6    Discussion

### 6.1    Interpretation of Simulation Results by Comparing with Experiments Involving Human Participants

The fact that there were no significant differences in the MSIs among the three tested motion profiles under the C_Control condition indicates that the motion strengths of the profiles were so similar to each other that the MSI differences could not be described by the conventional method. Under the C_GPR condition, in which our proposed method was used, a significantly smaller MSI was found under the M_P condition that yielded a more predictable motion profile compared with the other conditions, where in no significant difference was found. These tendencies agree with those observed in the human experiments conducted by Kuiper et al. (2019), in which almost the same motion profiles were used as those in the present study. This finding demonstrates that our proposed method with learning and prediction of motion dynamics can describe the effects of the predictability on the motion sickness subject to the condition that the motion strengths of the profiles were similar to each other, and the differences in the MSIs cannot be explained by the conventional method. Kuiper et al. (2019) were the first to investigate the effects of prediction on motion sickness systematically based on the separation of the predictability in motion direction and time. Few other systematic studies exist on this topic. A deeper understanding of the effects of predictions on the sickness based on additional experiments is essential to improve the accuracy of computational models. To the best of the author's knowledge, the present research is the first study that has quantitatively and systematically dealt with the effect of the predictability of exogenous motion on sensory conflict.

## 6.2    Comparison with Existing Modeling Research

Past experiences associated with the given motion stimulus and/or predictions of the future motion reduce sickness (Reason, 1978; Rolnick and Lubow, 1991). For the active participation to the body motion, the effect of motion prediction by the efference copy has progressed (Oman, 1982; Wada et al., 2018). However, the effects of the other types of prediction, such as the periodicity of motion patterns and visual information, have not been computationally understood at all. In fact, not only the models based on sensory conflict theory but also all the existing mathematical models, including fitting models, such as MSI (O'Hanlon and McCauley, 1974) and MSDV (ISO2631-1, 1997), have been unable to treat these effects. Hence, to the best of the author's knowledge, the present research was the first in which a conceptual model of the effects of knowledge of the exogenous motion on motion sickness has been built. The proposed model was consistently integrated with the conventional model describing the internal model hypothesis and effects to reduce the motion sickness caused by the efference copy (Fig. 2).

The second contribution of the present research is the development of a computational model to describe the effects of learning and prediction of the periodicity of motion on motion sickness. As a result, the present study demonstrates that the proposed single model can describe the tendency of sickness owing to the unpredictability of the motion direction and motion timing observed in the experiments conducted by Kuiper et al. (2019). Kuiper et al. (2020) and Karjanto et al. (2018) experimentally showed that the addition of artificial auditory and visual cues regarding the direction and/or timing of the motion in the future reduces motion sickness. The effects of such artificial cues are thought to be expressed by adding them as inputs to the proposed model. Further, the presence of forward vision reduces car sickness (Griffin and Newman, 2004; Wada and Yoshida, 2016). Forward vision has two effects: perception of current self-motion and prediction of future motion. For the former, computational models dealing with the visual–vestibular sensory interactions have been developed (Bos et al., 2007; Wada et al., 2020). The latter, which deals with future motion predictions, can be explained within the framework of Fig. 2, despite its differences from the perspective of motion dynamics learning as has been proposed in the present study (Fig. 3). Building a computational model to account for this effect is an important future task.

## 6.3    Relationship Between the Results and Neural Mismatch Theory

Neural mismatch theory (Reason, 1978) postulates that the sensory afferent is compared with the estimated signal obtained by the "neural store," in which history of the motion patterns is accumulated. This aspect has been modeled by the internal model hypothesis or the observer approach with the efference copy inputs (Oman, 1982; Wada et al., 2018). Neural mismatch theory (Reason, 1978) also states that "… the only strategy available to the neural store is to bias selection in favour of recently stored trace combinations," which constitute representations of another function or form of the neural store to decrease sickness by predicting exogenous motion when moved passively. The present study provides a computational implementation of this aspect of exogenous motion prediction. Note that although the proposed computational model is specialized for the SVC theory, the motion prediction block in the model can also be connected to the any other types of model that is based on observer theory such as sensory conflict theory (Oman, 1982) as shown in Fig.2.

## 6.4 Relationship with Strong and Weak Anticipation

It has been shown that complex systems exist in which there is synchronization of motions between multiple agents, and that the predictive motion can be achieved without having models of each other. This type of behavior is called strong anticipation (Dubois, 2003; Stepp and Turvey, 2010; Delignières and Marmelat, 2014). The concept of the strong anticipation implies that in human postural control systems, the dynamics of a closed-loop system of perception-action, including interactions with the environment, could have or include a function of motion prediction (Dubois, 2003). In contrast, explicit prediction of the future state of the target system based on its model is called weak anticipation. The prediction of future body state by the internal model dealt with in the present study can be regarded as weak anticipation. Furthermore, the block of exogenous motion prediction, which is expressed by RGPR, as proposed in this study can be also regarded as weak anticipation. The effect of the exogenous motion prediction in the context of the postural instability theory (Riccio and Stoffregen, 1991), which considers that motion sickness develops owing to inadequate postural control that occurs in the interaction between environment and body movement, could be understood using the strong anticipation concept as follows. Namely, if strong anticipation works effectively in the closed-loop dynamics of perception-action, including the interaction with the given environment, and it facilitates an improvement in postural control, it may reduce motion sickness. Prior literature wherein synchronization of visual information and postural variability was associated with motion sickness (Walter, et al., 2019) suggested a relationship between motion sickness and synchronization of vibration and body movement according to the strong anticipation concept. It should be noted that if the improvement of self-motion perception based on the weak anticipation concept leads to the improvement of postural control, the proposed model for exogenous motion prediction should be able to explain the effect of prediction on motion sickness in the context of postural instability theory in a broad sense, though the present research depends on sensory conflict theory.

## 6.5 Limitations and Future Research Questions

In the proposed model, RGPR was used for sequential learning and prediction of the exogenous motion. Methods that can learn time series data with strong nonlinearity and estimate the mean with confidence could serve as perspective candidates. Additional research of both the experimental and computational types is needed to deepen our understanding of the prediction effects and to improve the accuracy of the model. The rationale behind the selection of the free parameters and the function used to tune $k_{pr}$ in the proposed model should be investigated in the future because they were determined by trial and error. Given that the proposed model was evaluated only in a very limited number of scenarios, verification in various additional scenarios is needed. To verify the validity of the model and improve its accuracy, it is necessary to conduct a series of experiments on the effect of the prediction on motion sickness. Given the limited experimental knowledge of the effect of prediction on motion sickness, the proposed model is expected to facilitate the deriving directions or formulation of hypotheses for further experimental studies. For example, investigating the effect of period of the disturbance motion on motion sickness can reveal the size of the prediction time window, which will provide hints on the structure of the model used for human prediction.

The proposed model adjusts the degree of utilization of the predicted result for self-motion perception according to the variance of the predicted motion. This function was introduced based on an analogy with the multisensory integration process (Ernst, M.O and Bülthoff, 2004). This model function implies that the sickness will worsen when the difference between the estimated and actual values suddenly becomes large in the case that the variance of the prediction part is small, meaning that the confidence of the prediction is high. To the best of the author's knowledge, no prior research study has confirmed this implication experimentally. Thus, this topic also constitutes an interesting future research direction. The proposed computational model shown in Figure 3 can deal with only passive motions. Expansion to the model that can be applied even for active motions is also important future research. With the results of the present research, it is not possible to determine the extent to which the model reflects the underlying causal processes at present. It is expected that the interaction between hypothesis modeling research such as the present research and experimental research will advance this understanding.

## 7    Conflict of Interest

The author declares that a patent application on the model has been recently submitted by Ritsumeikan University.

## 8    Author Contributions

T.W. conceived of and designed the study, performed research, analyzed data, contributed new methods and models, and wrote the paper.

## 9    Funding

This work was partially supported by JSPS KAKENHI (Grant number 18H01414).

## 10    Acknowledgments

I wish to thank Dr. Yuki Okafuji for his advice on the computational method. This manuscript has been released as a preprint at arXiv (Wada, 2020).

**Tables**

Table 1. Model parameters

| $\tau$ [s] | $\tau_I$ [s] | K | b [m/s$^2$] | P [%] |
|:---:|:---:|:---:|:---:|:---:|
| 2.0 | 720 | 5.0 | 0.5 | 85.0 |